\documentclass[twocolumn]{aastex631}

\usepackage{graphicx}	% Including figure files
\usepackage{amsmath}	% Advanced maths commands

\begin{document}

\title{Can A Kinematically Hot and Thick Disk Form A Bar? : Role of Highly Spinning Dark Matter Halos}

\correspondingauthor{Sandeep Kumar Kataria}
\email{skkataria.iit@gmail.com}

\author[0000-0003-3657-0200]{Sandeep Kumar Kataria}
\affiliation{ Department of Space, Planetary \& Astronomical Sciences and Engineering, Indian Institute of Technology Kanpur, Kanpur 208016, India}

%% Note that the \and command from previous versions of AASTeX is now
%% depreciated in this version as it is no longer necessary. AASTeX 
%% automatically takes care of all commas and "and"s between authors names.

%% AASTeX 6.31 has the new \collaboration and \nocollaboration commands to
%% provide the collaboration status of a group of authors. These commands 
%% can be used either before or after the list of corresponding authors. The
%% argument for \collaboration is the collaboration identifier. Authors are
%% encouraged to surround collaboration identifiers with ()s. The 
%% \nocollaboration command takes no argument and exists to indicate that
%% the nearby authors are not part of surrounding collaborations.

%% Mark off the abstract in the ``abstract'' environment. 
\begin{abstract}
Recent JWST observations claim the existence of a significant fraction of bars in the kinematically hotter and thicker disk at high redshift Universe. These observations challenge the current understanding of disk stability in galaxies similar to the Milky Way. The analytical work and N-body simulations suggest that the kinematically hot (dispersion-dominated) and thick disk are stable against bar formation. In this work, we perform the controlled N-body simulations of a kinematically hot and thick disk, which is residing in a non-rotating and spinning dark matter halo. We report that the disk, which is classically stable against bar instability in the live and non-rotating halo, leads to bar formation in a spinning halo environment. The spinning halo model is 8 times more efficient in transporting angular momentum from the disk to the halo compared to the non-spinning halo. We claim that Ostriker-Peebles and ELN bar-formation criteria do not predict bar formation for both the non-roating and spinning halo. The recent criteria from Jang-Kim successfully predict the bar stability for the non-rotating halo model, but not for the spinning halo model. These results provide an important insight into the bar formation processes for thick and hot disks at high redshift. 

\end{abstract}

\section{Introduction} \label{sec:intro}
Observations claim that nearly two-thirds of disk galaxies in the nearby universe possess a bar \citep{Eskridge.et.al.2000,Sheth.et.al.2008,Erwin.2018}. The bar fraction is higher in early-type spiral galaxies (Sa-Sb) compared to the late type spirals (Sc-Sd) \citep{Lee.et.al.2019}. The fraction of barred galaxies at very high redshift has been debated \citep{Jogee.etal.2004, Sheth.et.al.2008}  though bar fraction is likely to increase with redshift as shown by the highly sensitive NIRCAM observations of JWST \citep{Guo.et.al.2023, Le-conte.et.al.2024,Guo.et.al.2024} recently. Despite widespread observations of barred galaxies, understanding the bar formation processes remain an open question.  

\cite{Toomre.1964} shows analytically that the gravitational instability of a rotationally supported disk forms a bar due to growing non-axisymmetric perturbations via swing amplification. The bar instability is triggered by internal perturbations \citep{Athanassoula.2003} or external perturbations \citep{Lokas.et.al.2014, Zheng.Shen.2025}, a thorough review is given in \cite{Sellwood.2014}. The galactic disk forms a bar as a result of angular momentum exchange to the surrounding dark-matter halo \citep{Debattista.Sellwood.2000,Athanassoula.2003}. Analytical calculations and numerical simulation studies of isolated disk evolution \citep{Ostriker.Peebles.1973, Kataria.Das.2018, Kataria.Das.2019, Saha.elmegreen.2018, Jang.Kim.2023, Chen_Shen.2025} show that dynamically cold disks are prone to bar instability, which is supported by observations \citep{Barazza.et.al.2008,Kataria_etal_2020}. The cosmological simulations also show that galaxies with central baryon domination prefer bars given a low bulge fraction in the galactic disk \citep{Rosas-Guevara.et.al.2020,Yetli.et.al.2022,Ansar.et.al.2023,Fragkoudi.et.al.2024, Kataria.Vivek.2024}. However, bar formation processes are significantly affected by the environments and gas fractions in the disks \citep{Yetli.et.al.2024,Ansar.et.al.2025}. 

One of the key parameters that affects the bar-triggering time scale is the central baryonic mass in the disk compared to the total galaxy mass \citep{Fujii.et.al.2019,Bland-Hawthorn.et.al.2023, Bland-Hawthorn.et.al.2024}. These studies claim that bar triggers earlier in the disks with higher fraction of baryonic content for a given kinematics and total mass in the galaxy. The other important parameter that affects the bar formation time scale is the spin of the dark-matter halo \citep{Kanak.Saha.Naab.2013,Collieretal.2018, Kataria.Shen.2022,Ansar.Das.2024,Kataria.Shen.2024,Chen_2025}. The transfer of angular momentum from disk to surrounding halo is efficient in rapidly rotating dark matter halo \citep{Chiba.Kataria.2024}.  These insights are helpful in understanding the recent observations \citep{Costantin.et.al.2023, Guo.et.al.2023, Le-conte.et.al.2024,Amvrosiadis.et.al.2025} of bar formation in the galaxies at early universe. Disks are very thick and turbulent or kinematically hotter at higher epochs \citep{Lian_2024}, which must be stable against bar formation in the classical regime \citep{Ostriker.Peebles.1973}. Therefore, it is necessary to understand the physical processes responsible for bar formation in kinematically hot and thick disks.   

In this article, we aim to explore the role of halo spin on the bar-forming modes of a thick and hot disk, which is bar stable a priori in a non-rotating halo. The presentation of the work follows as the numerical method in section \ref{Numerical section} and results in section \ref{Results}. Then, we discuss insights from the work in discussion section \ref{Discuss} and summarize the paper in section \ref{Summary}.

\section{Numerical Methods}\label{Numerical section} To understand the impact of halo spin on vertically thick and dispersion-dominated bar-stable disk in the non-spinning halo, we start with two models of galaxy initial conditions. First, we generate a Milky Way-like galaxy model (S000) using the AGAMA code \citep{Vasiliev.2019}. The total mass of the galaxy model is 5.49 $\times$ $10^{11} M_{\odot}$, which comprises a stellar disk and non-rotating dark matter halo components having a million particles each. 

\begin{figure}
    \centering
    \includegraphics[width=1 \linewidth]{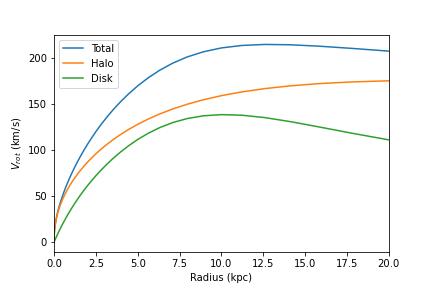}
    \caption{The galaxy model's rotation curve and the contribution from disk and halo separately.}
    \label{fig:rotcurve}
\end{figure}

\begin{figure}
    \centering
    \includegraphics[width=1 \linewidth]{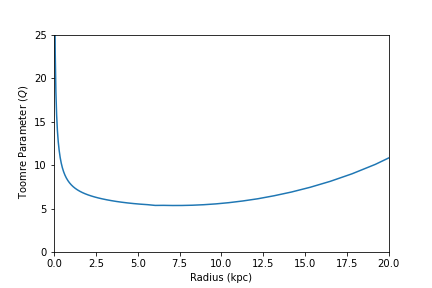}
    \caption{Radial variation of Toomre Q parameter for the initial galaxy disk for galaxy model.}
    \label{fig:ToomreQ}
\end{figure}

The following truncated Hernquist profile gives the dark matter halo density distribution.

\begin{equation}
    \rho_{DM}=\dfrac{M_{DM}}{2\pi} \dfrac{a}{r(r+a)^3} \times \exp[{-(r/r_{cut})^2}]
\end{equation}
Here $M_{DM}=5 \times 10^{11} M_{\cdot}$ is the dark matter halo mass, $a= 24 $ kpc and the cut-off radius of the Hernquist halo is 200 kpc.

The disk density distribution of the galaxy model follows the following profile.
\begin{equation}
    \Sigma = \dfrac{M_{\star}}{4 \pi R_d^{2}h} \exp{(R/R_d)} \text{sech}^2(z/h)
\end{equation}

Here $M_{\star}=4.9 \times 10^{10} M_{\cdot} $ is the mass of the stellar disk. $R_d=3.5$ kpc is an exponential profile and $h=0.7$ kpc is the height of the vertical scale.  The dispersion of the central disk on the disk is set to 175 Km/s. One of the local stability parameters is the Toomre ($Q$) which is calculated as ${\kappa}{\sigma}/(3.36G{\Sigma})$. Here $\kappa$ is the epicyclic frequency, ${\sigma}$ is the velocity dispersion, and ${\Sigma}$ is the stellar surface density. Figure \ref{fig:ToomreQ} shows the radial variation of the Toomre $Q$ parameter with $Q_{min}=5.35$ which is purposely chosen to be high value. This high value of Toomre $Q>>1$ ensures that the disk is classically stable against nonaxisymmetric modes \citep{BT.2008}. We have also checked the various classical criteria for bar formation listed in the literature. \\
\begin{itemize}
    \item{ \textbf{Ostriker and Peebles Criterion\citep{Ostriker.Peebles.1973}:} A disk is stable against bar formation if the ratio of rotational kinetic energy to total potential energy ($t_{OP}$) is less than 0.14 $\pm$ 0.003.}
    
    \item{\textbf{ELN Criterion \citep{Efstathiou.et.al.1982}}: A disk is bar unstable if 
    \begin{equation}
        \epsilon_{ELN}=\dfrac{V_{Max}}{\sqrt(M_{disc}G/R_d)} \ge 1.1
    \end{equation}
    here $V_{max}$ is the maximum velocity of galaxy rotation curve, $G$ is the gravitational constant, $R_d$ is the disk scale length and $M_{disc}$ is the stellar disk mass.}

    \item{\textbf{Jang and Kim Criterion ($t_{JK}$) \citep{Jang.Kim.2023}:} A disk is susceptible to bar formation if 
    \begin{equation}
        \bigg(\dfrac{Q_{min}}{1.2}\bigg)^2 +\bigg(\dfrac{CMC}{0.05}\bigg)^2 \le 1
    \end{equation}
    here, $CMC$ is the central mass concentration.}
\end{itemize}

\begin{center} 
\begin{table} 

    \begin{tabular}{|l|l|l|l|} 
         \hline
         Criterion &$t_{OP}$&$\epsilon_{ELN}$&$t_{JK}$  \\ \hline
         Values & 0.19 & 1.01 & 19.88 \\  \hline
    \end{tabular}    
    \caption{first row: $t_{OP}$, $\epsilon_{ELN}$ and $t_{JK}$ corresponds to Ostriker and Peebles criterion \citep{Ostriker.Peebles.1973}, ELN criterion \citep{Efstathiou.et.al.1982} and  Jang \& Kim criterion \citep{Jang.Kim.2023}, respectively. The values in the second row correspond to both models (S000 and S100).}
    \label{table}
\end{table}
\end{center}   
    %\item JK22

Table \ref{table} shows the values of all the three bar formation criteria discussed above. This shows that Ostriker-Peebles and ELN criteria predict bar formation for S000 model while the Jang-Kim criterion does not predict bar formation.

We calculate the spin of dark matter halo using the following definition \citep{Bullock.et.al.2001}:
    
    \begin{equation}
         \Lambda=\dfrac{J}{\sqrt{2GMR}}  
         \label{eq:lambda}
    \end{equation}
       
Here, $J$ is the magnitude of the specific angular momentum of the halo, $M$ is the mass of the halo within the virial radius, and $R$ is the virial radius of the halo. For the second galaxy model (S100), we increase the spin ($\Lambda$) of the dark matter halo to 0.1 using the methodology mentioned in \cite{Kataria.Shen.2024}. Hence, we ensure that the halo angular momentum distribution remains continuous while spinning the halo. The idea to choose spin ($\Lambda$=0.1) is to test for maximum impact of spin which is under the allowed range of spins in the cosmological simulations \citep{Bullock.et.al.2001}.  

We have evolved the galaxy models using the GADGET-2 code \citep{Springel.2005} until 9.78 Gyr. The angular momentum and energy conservation are within 0.1$\%$ in all the simulations. We use the softening length for the halo and disk components as 30 and 25 pc, respectively. We also perform the a test simulation by increasing the number of particles by a factor of 10 which shows that our results converge. We set the opening angle of the tree algorithm to be 0.4. Throughout the paper, we describe our results in terms of code units i.e. unit mass equal to $10^{10}$  M$_{\sun}$, unit velocity is 1 Km/s, and unit length is 1 kpc.

\section{Results} \label{Results} Given our interest in the growth of bar instability, we quantify the bar formation and evolution in the galactic disk using $m=2$ Fourier mode in the following manner \citep{Combes.Sanders.1981, Athanassoula.2003, Kataria.Das.2018}. 

\begin{equation}
a_2(R)=\sum_{i=1}^{N}  m_i \cos(2\phi_i)\\ \hspace{0.5cm}
b_2(R)=\sum_{i=1}^{N} m_i \sin(2 \phi_i)
\label{equation:FM}
\end{equation}

where $a_2$ and $b_2$ are calculated for the disk particles in concentric radial bins of 1 kpc throughout the disk, $m_i$ is the mass of $i^{th}$ star, $\phi_i$ is the azimuthal angle. We have defined the bar strength as the maximum value of normalised $A_2$ mode among all the concentric bins of 1 kpc size. 
\begin{equation}
\frac{A_2}{A_0}= max \Bigg(\frac{\sqrt{a_2 ^2 +b_2 ^2}}{\sum_{i=1}^{N} m_i} \Bigg)
\label{eq:barstrength}
\end{equation}

\begin{figure}
    \centering
    \includegraphics[width=1\linewidth]{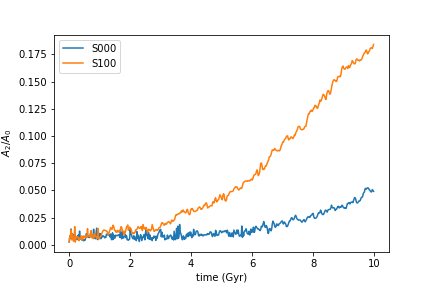}
    \caption{Time evolution of bar strengths for models with spin=0 (S000) and spin=0.1 (S100).}
    \label{fig:BS}
\end{figure}

\begin{figure*}
    \centering
    \includegraphics[width=1\linewidth]{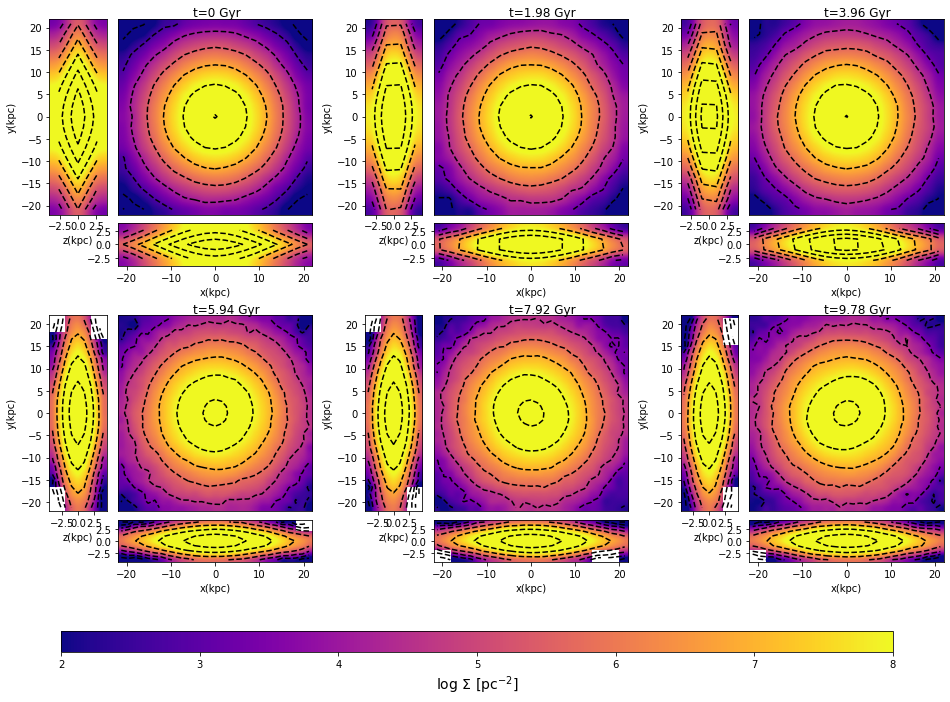}
    \caption{Time evolution of face-on and edge-on stellar density maps of disk for the model with spin=0 (S000). The dashed line corresponds to iso-density contours in each of the maps.}
    \label{fig:Faceon S000}
\end{figure*}

\begin{figure*}
    \centering
    \includegraphics[width=1\linewidth]{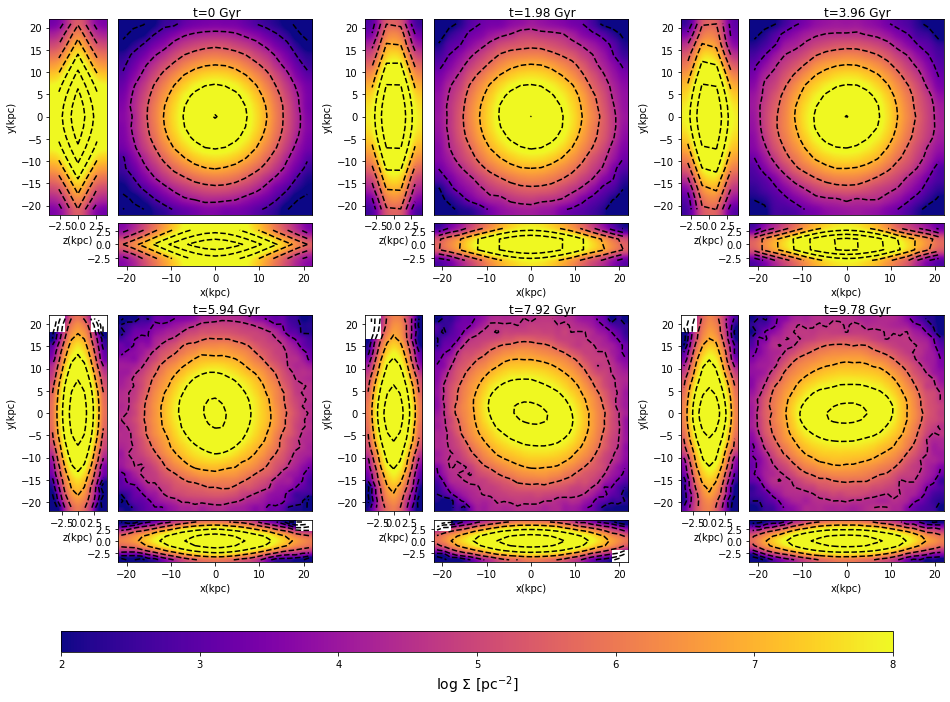}
    \caption{Time evolution of face-on and edge-on stellar density maps of disk for a model with spin=0.1 (S100). The dashed line corresponds to iso-density contours in each of the maps.}
    \label{fig:Faceon S100}
\end{figure*}

Figure \ref{fig:BS} shows the time evolution of simulated galaxy models. Model S000, having a non-rotating dark matter halo, does not allow bar formation until 10 Gyr.  We find that the disk remains axisymmetric throughout its evolution in the face-on stellar density maps, as shown in Figure \ref{fig:Faceon S000}. We find that Ostriker-Peebles ($t_{OP}$) and ELN ($\epsilon_{ELN}$) criteria do predict that disk should be unstable against bar formation while Jang-Kim ($t_{JK}$) criterion correctly predits stability of the disk against bar formation. On the other hand, the model S100 with the same disk kinematic while having a halo spin ($\Lambda$) of 0.1 shows the growth of bar instability at around 7 Gyr ($A_2/A_0>0.1$). The growth of $m=2$ Fourier mode is shown as the non-axisymmetric feature in stellar density maps of Figure \ref{fig:Faceon S100}. We find that all the three bar formation criteria namely Ostriker-Peebles, ELN, and Jang-Kim fail to predict bar formation in this model. 

\begin{figure}
    \centering
    \includegraphics[width=1\linewidth]{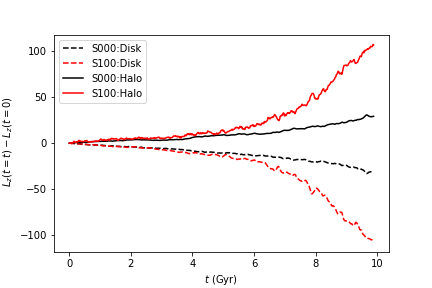}
    \caption{The total angular momentum exchange with time for disk and halo components of simulated galaxy models.}
    \label{fig:AM}
\end{figure}

To understand the dynamics of the simulated models, we plot the angular momentum of the disk and halo components in Figure \ref{fig:AM}. The disk loses angular momentum to the surrounding dark matter through resonance interactions \citep{Weinberg.1985, Chiba.2023, Chiba.Kataria.2024}. The amount of angular momentum lost by the disk is around 8 times higher in model S100 compared to S000. Therefore, we report that spinning dark matter halo can allow bar-type unstable modes in thick and dispersion-dominated disks. The effect of spin on bar forming mode is not being captured by the existing bar formation criteria in the literature.

\section{Discussion} \label{Discuss}

 The gist of the previous bar formation criteria ($t_{OP}$, $\epsilon_{ELN}$ and $t_{JK}$) suggest that a kinematically cold disk is susceptible to bar formation. However, the first two criteria ($t_{OP}$ and $\epsilon_{ELN}$) are limited by the rigid nature of the surrounding dark matter halo, which is not the case with $t_{JK}$. Therefore, for our non-spinning (but live) dark-matter halo model (S000), the $t_{OP}$ and $\epsilon_{ELN}$ criterion are not able to predict bar formation, which is not the case with the criterion $t_{JK}$. However, for the spinning-halo model (S100), all three criteria are unable to predict bar formation. Hence, this paper provides key insight into the predictable nature of the existing bar formation criteria.

Recent studies show that the transport of angular momentum from the disk to the surrounding dark matter halo enhances with the spin ($\Lambda$) of the halo \citep{Kanak.Saha.Naab.2013,Collieretal.2018,Kataria.Shen.2022,Chiba.Kataria.2024}. However, these studies focussed on the disk, which is a priori unstable to bar formation mode. Hence, the novelty of the current paper is to study the role of halo spin on the disk stable against bar formation, having very high value of Toomre $Q_{min}=5.35$. This study has implications for bar formation in kinematically hot and thick disks found at high redshift \citep{Lian_2024}. Previous studies show that thick and kinematically hot disks \citep{Zheng.Shen.2025} are stable against bar instability, while the simultaneous presence of an extra thin component promotes bar formation \citep{Ghosh.et.al.2023}.  Additionally, this study provides an explanation of the detection of bars at high redshift in kinematically hot and thick disks \citep{Guo.et.al.2023,Le-conte.et.al.2024,Amvrosiadis.et.al.2025}.  

 We find that the existing bar formation criteria \citep{Ostriker.Peebles.1973, Efstathiou.et.al.1982} are limited by the rigid nature of dark matter halo. while other criteria which include live dark matter halo \citep{Kataria.Das.2018,Saha.elmegreen.2018,Jang.Kim.2023}, also accounting for other properties like bulge mass, concentration and dark matter halo concentration etc.,  do not involve the role of dark matter halo spin, rather focusing on the dark matter potential and baryon properties only. Hence this study guides towards a bar formation criterion which also account for halo spin and better predictability power.

\section{Summary}\label{Summary}
We perform controlled N-body simulations of a dispersion-dominated and thick stellar disk within non-rotating and spinning dark matter halos respectively. We report that the stellar disk which is stable against bar-forming mode in the former case, becomes bar unstable for the highly spinning halo. Our key findings are

\begin{itemize}

    \item {The stellar disk within the non-rotating halo (S000) does not show bar-type instability within 10 Gyr of evolution. On the other hand, the same disk shows bar instability ($A_2/A_0 >0.1$) around 7 Gyr of evolution.} 
    \item{ The total angular momentum transport from the disk to the halo for the rotating halo model (S100) is enhanced by a factor of 8 compared to the non-rotating halo model (S000). This enhanced angular momentum exchange is crucial to the bar formation.}

    \item {The Ostriker-Peebles \& ELN ($\epsilon_{ELN}$) criteria values suggest that the non-rotating halo galaxy model (S000) is stable against the bar instability. The predictability of both the bar formation criteria is limited by the live nature of the dark matter halo.}

    \item{The Jang-Kim criterion ($t_{JK}$) predicts the stability against bar formation S000 model which fails for the spinning halo model (S100). Therefore, the spinning halo limits the predictable nature of the Jang-Kim criterion.}
    
\end{itemize}

The above results can potentially explain the bar formation in the dispersion-dominated and thick disks, which are considered stable against the bar formation. Recent observations of high redshift barred galaxies \citep{Guo.et.al.2023,Costantin.et.al.2023,Guo.et.al.2024,Le-conte.et.al.2024}, which are normally thicker and dispersion dominated, show a bar at high redshift. Our results suggests that observed bars at high redshift galaxies where the disks are thicker \citep{Lian_2024}, should possess highly rotating halos.
The physical processes underlying the angular momentum exchange in rotating halos are explained in these studies \citep{Weinberg.1985,Chiba.Kataria.2024}.

\begin{acknowledgments}

SKK acknowledges the support from the INSPIRE Faculty award (DST/INSPIRE/04/2023/000401) from the Department of Science and Technology, Government of India. We thank Juntai Shen and Yirui Zheng for the discussion related to the initial condition of the galaxy model. We thank Volker Springel for the GADGET code we used to run our simulations. This work used the Param Sanganak Supercomputing facility at IIT Kanpur, the Gravity Supercomputer at the Department of Astronomy, Shanghai Jiao Tong University, and the Centre for High-Performance Computing facilities at Shanghai Astronomical Observatory. 

\end{acknowledgments}

%% To help institutions obtain information on the effectiveness of their 
%% telescopes the AAS Journals has created a group of keywords for telescope 
%% facilities.
%
%% Following the acknowledgements section, use the following syntax and the
%% \facility{} or \facilities{} macros to list the keywords of facilities used 
%% in the research for the paper.  Each keyword is check against the master 
%% list during copy editing.  Individual instruments can be provided in 
%% parentheses, after the keyword, but they are not verified.

\vspace{7mm}

%% Similar to \facility{}, there is the optional \software command to allow 
%% authors a place to specify which programs were used during the creation of 
%% the manuscript. Authors should list each code and include either a
%% citation or url to the code inside ()s when available.

\software{numpy \citep{Harris.et.al.2020}, matplotlib \citep{Hunter.2007}, pynbody \citep{Pynbody.2013} and astropy \citep{2013A&A...558A..33A,2018AJ....156..123A}
          }

%% Appendix material should be preceded with a single \appendix command.
%% There should be a \section command for each appendix. Mark appendix
%% subsections with the same markup you use in the main body of the paper.

%% Each Appendix (indicated with \section) will be lettered A, B, C, etc.
%% The equation counter will reset when it encounters the \appendix
%% command and will number appendix equations (A1), (A2), etc. The
%% Figure and Table counter will not reset.

\bibliography{sample631}{}

@ARTICLE{2018AJ....156..123A,
       author = {{Astropy Collaboration} and {Price-Whelan}, A.~M. and {Sip{\H{o}}cz}, B.~M. and {G{\"u}nther}, H.~M. and {Lim}, P.~L. and {Crawford}, S.~M. and {Conseil}, S. and {Shupe}, D.~L. and {Craig}, M.~W. and {Dencheva}, N. and {Ginsburg}, A. and {VanderPlas}, J.~T. and {Bradley}, L.~D. and {P{\'e}rez-Su{\'a}rez}, D. and {de Val-Borro}, M. and {Aldcroft}, T.~L. and {Cruz}, K.~L. and {Robitaille}, T.~P. and {Tollerud}, E.~J. and {Ardelean}, C. and {Babej}, T. and {Bach}, Y.~P. and {Bachetti}, M. and {Bakanov}, A.~V. and {Bamford}, S.~P. and {Barentsen}, G. and {Barmby}, P. and {Baumbach}, A. and {Berry}, K.~L. and {Biscani}, F. and {Boquien}, M. and {Bostroem}, K.~A. and {Bouma}, L.~G. and {Brammer}, G.~B. and {Bray}, E.~M. and {Breytenbach}, H. and {Buddelmeijer}, H. and {Burke}, D.~J. and {Calderone}, G. and {Cano Rodr{\'\i}guez}, J.~L. and {Cara}, M. and {Cardoso}, J.~V.~M. and {Cheedella}, S. and {Copin}, Y. and {Corrales}, L. and {Crichton}, D. and {D'Avella}, D. and {Deil}, C. and {Depagne}, {\'E}. and {Dietrich}, J.~P. and {Donath}, A. and {Droettboom}, M. and {Earl}, N. and {Erben}, T. and {Fabbro}, S. and {Ferreira}, L.~A. and {Finethy}, T. and {Fox}, R.~T. and {Garrison}, L.~H. and {Gibbons}, S.~L.~J. and {Goldstein}, D.~A. and {Gommers}, R. and {Greco}, J.~P. and {Greenfield}, P. and {Groener}, A.~M. and {Grollier}, F. and {Hagen}, A. and {Hirst}, P. and {Homeier}, D. and {Horton}, A.~J. and {Hosseinzadeh}, G. and {Hu}, L. and {Hunkeler}, J.~S. and {Ivezi{\'c}}, {\v{Z}}. and {Jain}, A. and {Jenness}, T. and {Kanarek}, G. and {Kendrew}, S. and {Kern}, N.~S. and {Kerzendorf}, W.~E. and {Khvalko}, A. and {King}, J. and {Kirkby}, D. and {Kulkarni}, A.~M. and {Kumar}, A. and {Lee}, A. and {Lenz}, D. and {Littlefair}, S.~P. and {Ma}, Z. and {Macleod}, D.~M. and {Mastropietro}, M. and {McCully}, C. and {Montagnac}, S. and {Morris}, B.~M. and {Mueller}, M. and {Mumford}, S.~J. and {Muna}, D. and {Murphy}, N.~A. and {Nelson}, S. and {Nguyen}, G.~H. and {Ninan}, J.~P. and {N{\"o}the}, M. and {Ogaz}, S. and {Oh}, S. and {Parejko}, J.~K. and {Parley}, N. and {Pascual}, S. and {Patil}, R. and {Patil}, A.~A. and {Plunkett}, A.~L. and {Prochaska}, J.~X. and {Rastogi}, T. and {Reddy Janga}, V. and {Sabater}, J. and {Sakurikar}, P. and {Seifert}, M. and {Sherbert}, L.~E. and {Sherwood-Taylor}, H. and {Shih}, A.~Y. and {Sick}, J. and {Silbiger}, M.~T. and {Singanamalla}, S. and {Singer}, L.~P. and {Sladen}, P.~H. and {Sooley}, K.~A. and {Sornarajah}, S. and {Streicher}, O. and {Teuben}, P. and {Thomas}, S.~W. and {Tremblay}, G.~R. and {Turner}, J.~E.~H. and {Terr{\'o}n}, V. and {van Kerkwijk}, M.~H. and {de la Vega}, A. and {Watkins}, L.~L. and {Weaver}, B.~A. and {Whitmore}, J.~B. and {Woillez}, J. and {Zabalza}, V. and {Astropy Contributors}},
        title = "{The Astropy Project: Building an Open-science Project and Status of the v2.0 Core Package}",
      journal = {\aj},
         year = 2018,
        month = sep,
       volume = {156},
       number = {3},
          eid = {123},
        pages = {123},
          doi = {10.3847/1538-3881/aabc4f},
archivePrefix = {arXiv},
       eprint = {1801.02634},
 primaryClass = {astro-ph.IM},
       adsurl = {https://ui.adsabs.harvard.edu/abs/2018AJ....156..123A}
}

@article{Chen_2025,
doi = {10.3847/1538-4357/ae13a6},
url = {https://doi.org/10.3847/1538-4357/ae13a6},
year = {2025},
month = {nov},
publisher = {The American Astronomical Society},
volume = {994},
number = {1},
pages = {124},
author = {Chen, Bin-Hui and Kataria, Sandeep Kumar and Shen, Juntai and Guo, Meng},
title = {Dependency of the Bar Formation Timescale on the Halo Spin},
journal = {The Astrophysical Journal}
}

@ARTICLE{Ansar.et.al.2023,
       author = {{Ansar}, Sioree and {Kataria}, Sandeep Kumar and {Das}, Mousumi},
        title = "{Modelling dark matter halo spin using observations and simulations: application to UGC 5288}",
      journal = {\mnras},
         year = 2023,
        month = jun,
       volume = {522},
       number = {2},
        pages = {2967-2994},
          doi = {10.1093/mnras/stad1060},
archivePrefix = {arXiv},
       eprint = {2304.00724},
 primaryClass = {astro-ph.GA},
       adsurl = {https://ui.adsabs.harvard.edu/abs/2023MNRAS.522.2967A}
}

@ARTICLE{Chen_Shen.2025,
       author = {{Chen}, Bin-Hui and {Shen}, Juntai},
        title = "{The Dependency of Bar Formation Timescale on Disk Mass Fraction, Toomre Q, and Scale Height}",
      journal = {\apj},
         year = 2025,
        month = sep,
       volume = {990},
       number = {2},
          eid = {140},
        pages = {140},
          doi = {10.3847/1538-4357/adf966},
archivePrefix = {arXiv},
       eprint = {2510.13152},
 primaryClass = {astro-ph.GA},
       adsurl = {https://ui.adsabs.harvard.edu/abs/2025ApJ...990..140C}
}

@ARTICLE{2013A&A...558A..33A,
       author = {{Astropy Collaboration} and {Robitaille}, Thomas P. and
         {Tollerud}, Erik J. and {Greenfield}, Perry and {Droettboom}, Michael and
         {Bray}, Erik and {Aldcroft}, Tom and {Davis}, Matt and
         {Ginsburg}, Adam and {Price-Whelan}, Adrian M. and
         {Kerzendorf}, Wolfgang E. and {Conley}, Alexander and {Crighton}, Neil and
         {Barbary}, Kyle and {Muna}, Demitri and {Ferguson}, Henry and
         {Grollier}, Fr{\'e}d{\'e}ric and {Parikh}, Madhura M. and
         {Nair}, Prasanth H. and {Unther}, Hans M. and {Deil}, Christoph and
         {Woillez}, Julien and {Conseil}, Simon and {Kramer}, Roban and
         {Turner}, James E.~H. and {Singer}, Leo and {Fox}, Ryan and
         {Weaver}, Benjamin A. and {Zabalza}, Victor and {Edwards}, Zachary I. and
         {Azalee Bostroem}, K. and {Burke}, D.~J. and {Casey}, Andrew R. and
         {Crawford}, Steven M. and {Dencheva}, Nadia and {Ely}, Justin and
         {Jenness}, Tim and {Labrie}, Kathleen and {Lim}, Pey Lian and
         {Pierfederici}, Francesco and {Pontzen}, Andrew and {Ptak}, Andy and
         {Refsdal}, Brian and {Servillat}, Mathieu and {Streicher}, Ole},
        title = "{Astropy: A community Python package for astronomy}",
      journal = {\aap},
         year = "2013",
        month = "Oct",
       volume = {558},
          eid = {A33},
        pages = {A33},
          doi = {10.1051/0004-6361/201322068},
archivePrefix = {arXiv},
       eprint = {1307.6212},
 primaryClass = {astro-ph.IM},
       adsurl = {https://ui.adsabs.harvard.edu/abs/2013A&A...558A..33A}
}

@ARTICLE{Jang.Kim.2023,
       author = {{Jang}, Dajeong and {Kim}, Woong-Tae},
        title = "{Effects of the Central Mass Concentration on Bar Formation in Disk Galaxies}",
      journal = {\apj},
         year = 2023,
        month = jan,
       volume = {942},
       number = {2},
          eid = {106},
        pages = {106},
          doi = {10.3847/1538-4357/aca7bc},
archivePrefix = {arXiv},
       eprint = {2211.16816},
 primaryClass = {astro-ph.GA},
       adsurl = {https://ui.adsabs.harvard.edu/abs/2023ApJ...942..106J}
}

@ARTICLE{Bullock.et.al.2001,
       author = {{Bullock}, J.~S. and {Dekel}, A. and {Kolatt}, T.~S. and {Kravtsov}, A.~V. and {Klypin}, A.~A. and {Porciani}, C. and {Primack}, J.~R.},
        title = "{A Universal Angular Momentum Profile for Galactic Halos}",
      journal = {\apj},
         year = 2001,
        month = jul,
       volume = {555},
       number = {1},
        pages = {240-257},
          doi = {10.1086/321477},
archivePrefix = {arXiv},
       eprint = {astro-ph/0011001},
 primaryClass = {astro-ph},
       adsurl = {https://ui.adsabs.harvard.edu/abs/2001ApJ...555..240B}
}

@ARTICLE{Kataria_etal_2020,
       author = {{Kataria}, Sandeep Kumar and {Das}, Mousumi and {Barway}, Sudhanshu},
        title = "{Testing a theoretical prediction for bar formation in galaxies with bulges}",
      journal = {\aap},
         year = 2020,
        month = aug,
       volume = {640},
          eid = {A14},
        pages = {A14},
          doi = {10.1051/0004-6361/202037527},
archivePrefix = {arXiv},
       eprint = {2006.05870},
 primaryClass = {astro-ph.GA},
       adsurl = {https://ui.adsabs.harvard.edu/abs/2020A&A...640A..14K}
}

@BOOK{BT.2008,
       author = {{Binney}, James and {Tremaine}, Scott},
        title = "{Galactic Dynamics: Second Edition}",
         year = 2008,
    publisher = {Princeton University Press},
       adsurl = {https://ui.adsabs.harvard.edu/abs/2008gady.book.....B}
}

@ARTICLE{Barazza.et.al.2008,
       author = {{Barazza}, Fabio D. and {Jogee}, Shardha and {Marinova}, Irina},
        title = "{Bars in Disk-dominated and Bulge-dominated Galaxies at z \raisebox{-0.5ex}\textasciitilde 0: New Insights from \raisebox{-0.5ex}\textasciitilde3600 SDSS Galaxies}",
      journal = {\apj},
         year = 2008,
        month = mar,
       volume = {675},
       number = {2},
        pages = {1194-1212},
          doi = {10.1086/526510},
archivePrefix = {arXiv},
       eprint = {0710.4674},
 primaryClass = {astro-ph},
       adsurl = {https://ui.adsabs.harvard.edu/abs/2008ApJ...675.1194B}
}

@ARTICLE{Debattista.Sellwood.2000,
       author = {{Debattista}, Victor P. and {Sellwood}, J.~A.},
        title = "{Constraints from Dynamical Friction on the Dark Matter Content of Barred Galaxies}",
      journal = {\apj},
         year = 2000,
        month = nov,
       volume = {543},
       number = {2},
        pages = {704-721},
          doi = {10.1086/317148},
archivePrefix = {arXiv},
       eprint = {astro-ph/0006275},
 primaryClass = {astro-ph},
       adsurl = {https://ui.adsabs.harvard.edu/abs/2000ApJ...543..704D}
}

@ARTICLE{Eskridge.et.al.2000,
       author = {{Eskridge}, Paul B. and {Frogel}, Jay A. and {Pogge}, Richard W. and {Quillen}, Alice C. and {Davies}, Roger L. and {DePoy}, D.~L. and {Houdashelt}, Mark L. and {Kuchinski}, Leslie E. and {Ram{\'\i}rez}, Solange V. and {Sellgren}, K. and {Terndrup}, Donald M. and {Tiede}, Glenn P.},
        title = "{The Frequency of Barred Spiral Galaxies in the Near-Infrared}",
      journal = {\aj},
         year = 2000,
        month = feb,
       volume = {119},
       number = {2},
        pages = {536-544},
          doi = {10.1086/301203},
archivePrefix = {arXiv},
       eprint = {astro-ph/9910479},
 primaryClass = {astro-ph},
       adsurl = {https://ui.adsabs.harvard.edu/abs/2000AJ....119..536E}
}

@ARTICLE{Erwin.2018,
       author = {{Erwin}, Peter},
        title = "{The dependence of bar frequency on galaxy mass, colour, and gas content - and angular resolution - in the local universe}",
      journal = {\mnras},
         year = 2018,
        month = mar,
       volume = {474},
       number = {4},
        pages = {5372-5392},
          doi = {10.1093/mnras/stx3117},
archivePrefix = {arXiv},
       eprint = {1711.04867},
 primaryClass = {astro-ph.GA},
       adsurl = {https://ui.adsabs.harvard.edu/abs/2018MNRAS.474.5372E}
}

@ARTICLE{Lee.et.al.2019,
       author = {{Lee}, Yun Hee and {Ann}, Hong Bae and {Park}, Myeong-Gu},
        title = "{Bar Fraction in Early- and Late-type Spirals}",
      journal = {\apj},
         year = 2019,
        month = feb,
       volume = {872},
       number = {1},
          eid = {97},
        pages = {97},
          doi = {10.3847/1538-4357/ab0024},
archivePrefix = {arXiv},
       eprint = {1901.05183},
 primaryClass = {astro-ph.GA},
       adsurl = {https://ui.adsabs.harvard.edu/abs/2019ApJ...872...97L}
}

@ARTICLE{Jogee.etal.2004,
       author = {{Jogee}, Shardha and {Barazza}, Fabio D. and {Rix}, Hans-Walter and {Shlosman}, Isaac and {Barden}, Marco and {Wolf}, Christian and {Davies}, James and {Heyer}, Inge and {Beckwith}, Steven V.~W. and {Bell}, Eric F. and {Borch}, Andrea and {Caldwell}, John A.~R. and {Conselice}, Christopher J. and {Dahlen}, Tomas and {H{\"a}ussler}, Boris and {Heymans}, Catherine and {Jahnke}, Knud and {Knapen}, Johan H. and {Laine}, Seppo and {Lubell}, Gabriel M. and {Mobasher}, Bahram and {McIntosh}, Daniel H. and {Meisenheimer}, Klaus and {Peng}, Chien Y. and {Ravindranath}, Swara and {Sanchez}, Sebastian F. and {Somerville}, Rachel S. and {Wisotzki}, Lutz},
        title = "{Bar Evolution over the Last 8 Billion Years: A Constant Fraction of Strong Bars in the GEMS Survey}",
      journal = {\apjl},
         year = 2004,
        month = nov,
       volume = {615},
       number = {2},
        pages = {L105-L108},
          doi = {10.1086/426138},
archivePrefix = {arXiv},
       eprint = {astro-ph/0408382},
 primaryClass = {astro-ph},
       adsurl = {https://ui.adsabs.harvard.edu/abs/2004ApJ...615L.105J}
}

@ARTICLE{Guo.et.al.2023,
       author = {{Guo}, Yuchen and {Jogee}, Shardha and {Finkelstein}, Steven L. and {Chen}, Zilei and {Wise}, Eden and {Bagley}, Micaela B. and {Barro}, Guillermo and {Wuyts}, Stijn and {Kocevski}, Dale D. and {Kartaltepe}, Jeyhan S. and {McGrath}, Elizabeth J. and {Ferguson}, Henry C. and {Mobasher}, Bahram and {Giavalisco}, Mauro and {Lucas}, Ray A. and {Zavala}, Jorge A. and {Lotz}, Jennifer M. and {Grogin}, Norman A. and {Huertas-Company}, Marc and {Vega-Ferrero}, Jes{\'u}s and {Hathi}, Nimish P. and {Haro}, Pablo Arrabal and {Dickinson}, Mark and {Koekemoer}, Anton M. and {Papovich}, Casey and {Pirzkal}, Nor and {Yung}, L.~Y. Aaron and {Backhaus}, Bren E. and {Bell}, Eric F. and {Calabr{\`o}}, Antonello and {Cleri}, Nikko J. and {Coogan}, Rosemary T. and {Cooper}, M.~C. and {Costantin}, Luca and {Croton}, Darren and {Davis}, Kelcey and {Dekel}, Avishai and {Franco}, Maximilien and {Gardner}, Jonathan P. and {Holwerda}, Benne W. and {Hutchison}, Taylor A. and {Pandya}, Viraj and {P{\'e}rez-Gonz{\'a}lez}, Pablo G. and {Ravindranath}, Swara and {Rose}, Caitlin and {Trump}, Jonathan R. and {de la Vega}, Alexander and {Wang}, Weichen},
        title = "{First Look at z > 1 Bars in the Rest-frame Near-infrared with JWST Early CEERS Imaging}",
      journal = {\apjl},
         year = 2023,
        month = mar,
       volume = {945},
       number = {1},
          eid = {L10},
        pages = {L10},
          doi = {10.3847/2041-8213/acacfb},
archivePrefix = {arXiv},
       eprint = {2210.08658},
 primaryClass = {astro-ph.GA},
       adsurl = {https://ui.adsabs.harvard.edu/abs/2023ApJ...945L..10G}
}

@ARTICLE{Le-conte.et.al.2024,
       author = {{Le Conte}, Zoe A. and {Gadotti}, Dimitri A. and {Ferreira}, Leonardo and {Conselice}, Christopher J. and {de S{\'a}-Freitas}, Camila and {Kim}, Taehyun and {Neumann}, Justus and {Fragkoudi}, Francesca and {Athanassoula}, E. and {Adams}, Nathan J.},
        title = "{A JWST investigation into the bar fraction at redshifts 1 {\ensuremath{\leq}} z {\ensuremath{\leq}} 3}",
      journal = {\mnras},
         year = 2024,
        month = may,
       volume = {530},
       number = {2},
        pages = {1984-2000},
          doi = {10.1093/mnras/stae921},
archivePrefix = {arXiv},
       eprint = {2309.10038},
 primaryClass = {astro-ph.GA},
       adsurl = {https://ui.adsabs.harvard.edu/abs/2024MNRAS.530.1984L}
}

@ARTICLE{Toomre.1964,
       author = {{Toomre}, A.},
        title = "{On the gravitational stability of a disk of stars.}",
      journal = {\apj},
         year = 1964,
        month = may,
       volume = {139},
        pages = {1217-1238},
          doi = {10.1086/147861},
       adsurl = {https://ui.adsabs.harvard.edu/abs/1964ApJ...139.1217T}
}

@ARTICLE{Lokas.et.al.2014,
       author = {{{\L}okas}, E.~L. and {Athanassoula}, E. and {Debattista}, V.~P. and {Valluri}, M. and {Pino}, A. del and {Semczuk}, M. and {Gajda}, G. and {Kowalczyk}, K.},
        title = "{Adventures of a tidally induced bar}",
      journal = {\mnras},
         year = 2014,
        month = dec,
       volume = {445},
       number = {2},
        pages = {1339-1350},
          doi = {10.1093/mnras/stu1846},
archivePrefix = {arXiv},
       eprint = {1404.1211},
 primaryClass = {astro-ph.GA},
       adsurl = {https://ui.adsabs.harvard.edu/abs/2014MNRAS.445.1339L}
}

@ARTICLE{Zheng.Shen.2025,
       author = {{Zheng}, Yirui and {Shen}, Juntai},
        title = "{Comparison of Bar Formation Mechanisms. I. Does a Tidally Induced Bar Rotate Slower than an Internally Induced Bar?}",
      journal = {\apj},
         year = 2025,
        month = jan,
       volume = {979},
       number = {1},
          eid = {60},
        pages = {60},
          doi = {10.3847/1538-4357/ad9bae},
archivePrefix = {arXiv},
       eprint = {2412.04770},
 primaryClass = {astro-ph.GA},
       adsurl = {https://ui.adsabs.harvard.edu/abs/2025ApJ...979...60Z}
}

@ARTICLE{Sellwood.2014,
       author = {{Sellwood}, J.~A.},
        title = "{Secular evolution in disk galaxies}",
      journal = {Reviews of Modern Physics},
         year = 2014,
        month = jan,
       volume = {86},
       number = {1},
        pages = {1-46},
          doi = {10.1103/RevModPhys.86.1},
archivePrefix = {arXiv},
       eprint = {1310.0403},
 primaryClass = {astro-ph.GA},
       adsurl = {https://ui.adsabs.harvard.edu/abs/2014RvMP...86....1S}
}

@ARTICLE{Kataria.Das.2018,
       author = {{Kataria}, Sandeep Kumar and {Das}, Mousumi},
        title = "{A study of the effect of bulges on bar formation in disc galaxies}",
      journal = {\mnras},
         year = 2018,
        month = apr,
       volume = {475},
       number = {2},
        pages = {1653-1664},
          doi = {10.1093/mnras/stx3279},
       adsurl = {https://ui.adsabs.harvard.edu/abs/2018MNRAS.475.1653K}
}

@ARTICLE{Kataria.Shen.2022,
       author = {{Kataria}, Sandeep Kumar and {Shen}, Juntai},
        title = "{Effects of Inner Halo Angular Momentum on the Peanut/X Shapes of Bars}",
      journal = {\apj},
         year = 2022,
        month = dec,
       volume = {940},
       number = {2},
          eid = {175},
        pages = {175},
          doi = {10.3847/1538-4357/ac9df1},
archivePrefix = {arXiv},
       eprint = {2210.14526},
 primaryClass = {astro-ph.GA},
       adsurl = {https://ui.adsabs.harvard.edu/abs/2022ApJ...940..175K}
}

@article{Saha.elmegreen.2018,
doi = {10.3847/1538-4357/aabacd},
url = {https://dx.doi.org/10.3847/1538-4357/aabacd},
year = {2018},
month = {apr},
publisher = {The American Astronomical Society},
volume = {858},
number = {1},
pages = {24},
author = {Kanak Saha and Bruce Elmegreen},
title = {Why Are Some Galaxies Not Barred?},
journal = {The Astrophysical Journal}
}

@ARTICLE{Kanak.Saha.Naab.2013,
       author = {{Saha}, Kanak and {Naab}, Thorsten},
        title = "{Spinning dark matter haloes promote bar formation}",
      journal = {\mnras},
         year = 2013,
        month = sep,
       volume = {434},
       number = {2},
        pages = {1287-1299},
          doi = {10.1093/mnras/stt1088},
archivePrefix = {arXiv},
       eprint = {1304.1667},
 primaryClass = {astro-ph.CO},
       adsurl = {https://ui.adsabs.harvard.edu/abs/2013MNRAS.434.1287S}
}

@ARTICLE{Collieretal.2018,
       author = {{Collier}, Angela and {Shlosman}, Isaac and {Heller}, Clayton},
        title = "{What makes the family of barred disc galaxies so rich: damping stellar bars in spinning haloes}",
      journal = {\mnras},
         year = 2018,
        month = may,
       volume = {476},
       number = {1},
        pages = {1331-1344},
          doi = {10.1093/mnras/sty270},
archivePrefix = {arXiv},
       eprint = {1712.02802},
 primaryClass = {astro-ph.GA},
       adsurl = {https://ui.adsabs.harvard.edu/abs/2018MNRAS.476.1331C}
}

@ARTICLE{Yetli.et.al.2022,
       author = {{Rosas-Guevara}, Yetli and {Bonoli}, Silvia and {Dotti}, Massimo and {Izquierdo-Villalba}, David and {Lupi}, Alessandro and {Zana}, Tommaso and {Bonetti}, Matteo and {Nelson}, Dylan and {Springel}, Volker and {Hernquist}, Lars and {Vogelsberger}, Mark},
        title = "{The evolution of the barred galaxy population in the TNG50 simulation}",
      journal = {\mnras},
         year = 2022,
        month = jun,
       volume = {512},
       number = {4},
        pages = {5339-5357},
          doi = {10.1093/mnras/stac816},
archivePrefix = {arXiv},
       eprint = {2110.04537},
 primaryClass = {astro-ph.GA},
       adsurl = {https://ui.adsabs.harvard.edu/abs/2022MNRAS.512.5339R}
}

@ARTICLE{Rosas-Guevara.et.al.2020,
       author = {{Rosas-Guevara}, Yetli and {Bonoli}, Silvia and {Dotti}, Massimo and {Zana}, Tommaso and {Nelson}, Dylan and {Pillepich}, Annalisa and {Ho}, Luis C. and {Izquierdo-Villalba}, David and {Hernquist}, Lars and {Pakmor}, R{\"u}ediger},
        title = "{The buildup of strongly barred galaxies in the TNG100 simulation}",
      journal = {\mnras},
         year = 2020,
        month = jan,
       volume = {491},
       number = {2},
        pages = {2547-2564},
          doi = {10.1093/mnras/stz3180},
archivePrefix = {arXiv},
       eprint = {1908.00547},
 primaryClass = {astro-ph.GA},
       adsurl = {https://ui.adsabs.harvard.edu/abs/2020MNRAS.491.2547R}
}

@ARTICLE{Fragkoudi.et.al.2024,
       author = {{Fragkoudi}, Francesca and {Grand}, Robert and {Pakmor}, R{\"u}diger and {G{\'o}mez}, Facundo and {Marinacci}, Federico and {Springel}, Volker},
        title = "{Bar formation and evolution in the cosmological context: Inputs from the Auriga simulations}",
      journal = {arXiv e-prints},
         year = 2024,
        month = jun,
          eid = {arXiv:2406.09453},
        pages = {arXiv:2406.09453},
          doi = {10.48550/arXiv.2406.09453},
archivePrefix = {arXiv},
       eprint = {2406.09453},
 primaryClass = {astro-ph.GA},
       adsurl = {https://ui.adsabs.harvard.edu/abs/2024arXiv240609453F}
}

@ARTICLE{Amvrosiadis.et.al.2025,
       author = {{Amvrosiadis}, A. and {Lange}, S. and {Nightingale}, J.~W. and {He}, Q. and {Frenk}, C.~S. and {Oman}, K.~A. and {Smail}, I. and {Swinbank}, A.~M. and {Fragkoudi}, F. and {Gadotti}, D.~A. and {Cole}, S. and {Borsato}, E. and {Robertson}, A. and {Massey}, R. and {Cao}, X. and {Li}, R.},
        title = "{The onset of bar formation in a massive galaxy at z \raisebox{-0.5ex}\textasciitilde 3.8}",
      journal = {\mnras},
         year = 2025,
        month = feb,
       volume = {537},
       number = {2},
        pages = {1163-1181},
          doi = {10.1093/mnras/staf048},
archivePrefix = {arXiv},
       eprint = {2404.01918},
 primaryClass = {astro-ph.GA},
       adsurl = {https://ui.adsabs.harvard.edu/abs/2025MNRAS.537.1163A}
}

@ARTICLE{Fujii.et.al.2019,
       author = {{Fujii}, M.~S. and {B{\'e}dorf}, J. and {Baba}, J. and {Portegies Zwart}, S.},
        title = "{Modelling the Milky Way as a dry Galaxy}",
      journal = {\mnras},
         year = 2019,
        month = jan,
       volume = {482},
       number = {2},
        pages = {1983-2015},
          doi = {10.1093/mnras/sty2747},
archivePrefix = {arXiv},
       eprint = {1807.10019},
 primaryClass = {astro-ph.GA},
       adsurl = {https://ui.adsabs.harvard.edu/abs/2019MNRAS.482.1983F}
}
\bibliographystyle{aasjournal}

%% This command is needed to show the entire author+affiliation list when
%% the collaboration and author truncation commands are used.  It has to
%% go at the end of the manuscript.
%\allauthors

%% Include this line if you are using the \added, \replaced, \deleted
%% commands to see a summary list of all changes at the end of the article.
%\listofchanges

\end{document}